\documentclass[runningheads]{llncs}
\usepackage[T1]{fontenc}
\usepackage{float}
\usepackage{graphicx}
\usepackage[subfolder]{gnuplottex}
\usepackage{tikz}
\usepackage{subfigure}
\usepackage{caption}
\usepackage{subcaption}
\usepackage{setspace}
 \renewcommand{\abstractname}{Executive Summary}

\begin{document}

%
%
%
%
\title{Energy efficiency of cache eviction algorithms for Zipf distributed objects}
%
%
\author{Emese Sziklay \inst{1}\orcidID{0009-0002-3159-0202} \and \\
Tamás Jursonovics\inst{2}\orcidID{0000-0002-0804-5399} }
\authorrunning{E. Sziklay et al.}
%
\institute{Data Science and Engineering Department,
Faculty of Informatics, Eötvös Loránd University\\
Pázmány Péter str. 1/A, 1117 
Budapest, Hungary \\
\email{sziklayemese@inf.elte.hu}
\and
Deutsche Telekom, Deutsche-Telekom-Allee 9\\
64295 Darmstadt, Germany\\
\email{tamas.jursonovics@telekom.de}\\ }
\maketitle    

\begin{abstract}
    
This paper presents a summary analysis of the Least Frequently Used (LFU) and Perfect Least Frequently Used (PLFU) cache eviction algorithms on real data, transferred on Content Delivery Networks (CDNs), as well as on Zipf's distributed samples. In light of the growing emphasis on energy efficiency in CDNs in recent years due to rising energy costs, this paper considers and discusses the total CPU time required to run a cache algorithm. The total CPU time represents a novel metric for evaluating cache performance, and it is contrasted with the conventional Cache Hit Ratio (CHR) metric. Furthermore, a new algorithm with an admission policy and the eviction strategy that of PLFU is presented. The results demonstrate that it is a simple and  straightforward algorithm to implement and offers high CHR and low CPU time.
\end{abstract}
\section{Introduction}\label{introduction}

The advent of global streaming platforms such as Disney, Hulu or Netflix has elevated the significance of Content Delivery Networks (CDNs) to a degree that was previously unparalleled. The on-demand consumption of video and media content requires the implementation of scalable content delivery networks (CDNs) with sophisticated protocols and algorithms \cite{Thomdapu2021}. In light of the rising electricity prices across Europe since 2022, as reported by Eurostat \cite{EUROstat}, energy saving has become not only a general policy in the industry these days but also a key point of concern that could provide a competitive advantage for companies.
Although previous studies have investigated energy-efficient solutions for cache systems, proposing a flexible and configurable cache architecture \cite{sundararajan2013smart}, the efficiency of cache management in terms of the required total CPU time represents a simple but novel perspective.

This paper focuses on the performance metrics of cache management algorithms commonly employed in CDN architectures. Additionally, we propose a novel method, designated as Perfect Least Frequently Used with Admission (PLFUA) which leverages the access patterns of web objects in conjunction with the eviction strategy of PLFU. Our findings illustrate that PLFUA is a straightforward implementation, offers high CHR with minimal total CPU time and reduced overhead compared to LFU  or PLFU.

\subsection{LFU and PLFU cache eviction algorithms and cache performance metrics}

The literature contains numerous papers analyzing frequency-based cache eviction algorithms. For example, see \cite{Breslau1998},\cite{Breslau1999}. For further details, please refer to the following sources: \cite{Karakostas2000},\cite{Karakostas2002},\cite{Megiddo2004},\cite{Shi2005},\cite{Nair2010},\cite{Hasslinger2014} and \cite{Einziger2017}.
The Least Frequently Used (LFU) algorithm represents one eviction strategy among numerous others. Its objective is to optimize the management of cache storage, whereby the least frequently used objects are evicted when the cache is full and free space is needed for new objects. As stated in reference \cite{Karakostas2000}, the majority of web caches utilize the LFU replacement policy. Although the Least Recently Used (LRU) eviction strategy does not necessitate the use of any metadata and is relatively straightforward to implement, as demonstrated by~\cite{Ntougias2021}, its hit rate is typically low when applied to web objects. This is due to the fact that relevant information -- for predicting the next object to be requested -- is not exploited. For this reason, the two most commonly used LFU algorithms (in-memory LFU or LFU and perfect LFU, PLFU) are discussed of caching web data.

In order to implement the eviction strategy for LFU, it is necessary to create a single storage unit, or container, which will record the request frequencies of the cached elements. A request is defined as a hit if the specific content is present in the cache, otherwise it is classified as a miss. One of the most crucial aspects of any cache strategy is to ensure that the number of hits remains as high as possible, in other words, achieve high cache hit ratio (CHR). 

Perfect LFU, (PLFU) necessitates the utilization of two distinct containers during its operation. The initial container serves the function of storing the request frequencies of objects within the cache, in a manner analogous to that observed in LFU. The second container is employed to maintain a historical record of the frequencies associated with evicted cache objects. The additional metadata allows the frequency of the object to be used from the most recent recorded value, rather than starting from one as would be the case with LFU. In general, the space required for the eviction strategy is referred to as the metadata for the caching algorithm. In order to overcome the memory limitation problem of PLFU, a limited version (Window-LFU~\cite{Karakostas2000}) of this algorithm is used in practice, given that the container of historical data could grow indefinitely by new content requests.

It is crucial to highlight that the CHR is not the sole cache performance metric to take into account when designing a cache system. In \cite{Nair2010} and \cite{nagaraj2006web} the volume of bytes ingested (throughput), while in \cite{Ntougias2021} a detailed performance comparison of caching strategies is presented in terms of computation efforts per request, hit rate and information base. In \cite{Hasslinger2016}, the simulation methods used to evaluate caching efficiency are discussed. In reference \cite{nagaraj2006web} the cache age and the downtime are also considered as parameters for measuring performance.

In addition, the load on the CPU, and consequently the energy required for a cache algorithm, can also be considered a performance metric, thus representing an important factor for decision makers. In certain circumstances, the energy consumed by an algorithm can be quantified in Joules with the assistance of the Intel “Running Average Power Limit” (RAPL) technology, which estimates the energy expenditure of the CPU, RAM and integrated GPU.

The objective of future work will be to test cache algorithms with the RAPL technology. In this paper, the total CPU running time will be presented and used as an indicator to compare cache eviction strategies in terms of energy efficiency. This enables a direct comparison of different implementations without consideration of the CPU's own power-saving features (e.g., idle cores, P states.) While the total CPU execution time has not been previously discussed in the literature with regard to cache algorithms, it has been determined for eight sorting algorithms in \cite{Zeebaree2015} and for matrix multiplication in \cite{ElEnin2011}.

The algorithms employed in \cite{Rashid2018} are capable of calculating the total CPU execution time. However, it is also observed that the CPU-usage is influenced by under-running tasks, contingent on the  status of the computer. Furthermore, the discrepancies in the characteristics of the utilized hosts at the server level may yield disparate CPU usage values across different hosts, potentially leading to unforeseen outcomes in specific instances.

The total CPU time for cache algorithms was measured in this study using a Jupyter Lab on a physical server with an Intel® Xeon® Gold 6130 processor, 32 physical and 64 logical cores. In order to mitigate the impact of under-running tasks on the CPU, each caching algorithm was evaluated on 12 samples of 100,000, and the mean CPU times were calculated. The samples follow Zipf distribution with parameter 1.1, which is the best approximation to the real data obtained from a leading European Internet Service Provider.
The efficacy of three cache algorithms, namely LFU, PLFU and an own algorithm named PLFUA outperforms the other algorithms in terms of CPU time and CHR. The algorithms have been optimized and implemented in the same manner, ensuring that their performance in terms of CPU time is comparable.

\subsection{Review of frequency based algorithms in the literature}

The subject of cache eviction or replacement algorithms for media and web objects is a widely discussed topic in the literature. Some alternative implementations of LFU integrate the frequency and recency of cache objects, thereby creating a hybrid eviction strategy. 

Window-LFU, (WLFU) method, as described in \cite{Karakostas2000} and \cite{Karakostas2002} is a replacement decision-making process based on the frequency of access in a recent past period, referred to as a time window. In other words, the WLFU maintains the access frequency for a window of the last W requests \cite{Karakostas2002}. Nevertheless, this algorithm necessitates the monitoring of the sequence of requests, rendering it more adaptable to evolving circumstances than PLFU.

The replacement strategy identified in \cite{Nair2010} employs a caching system for video objects, wherein not only access frequency and recency are considered, but also other salient characteristics, including the size and cost of transfer of the video objects in question.
The aging method and geometric fading are discussed in detail in \cite{Hasslinger2014} and hit rates of the algorithms are compared. 

In the overwhelming of cache algorithms, a newly accessed item is always inserted into the cache with the caching strategy focusing exclusively on its eviction policy. In other words, the decision as to which item should be replaced is the sole aspect of the strategy that is of concern. In contrast, the Tiny-LFU and Window-Tiny-LFU employ an admission policy based on the frequencies of recently  accessed items. In order to maintain low metadata costs, a Bloom filter-style structure is employed, which provides an accurate approximation of the frequencies at a low cost while allowing for changes in access patterns over time \cite{Einziger2017}.

\section{PLFU is preferred to LFU for caching media data}\label{PLFU}

\subsection{LFU algorithm on media objects}

\begin{figure}[htp] 
\centering
\includegraphics[width=0.48\textwidth]{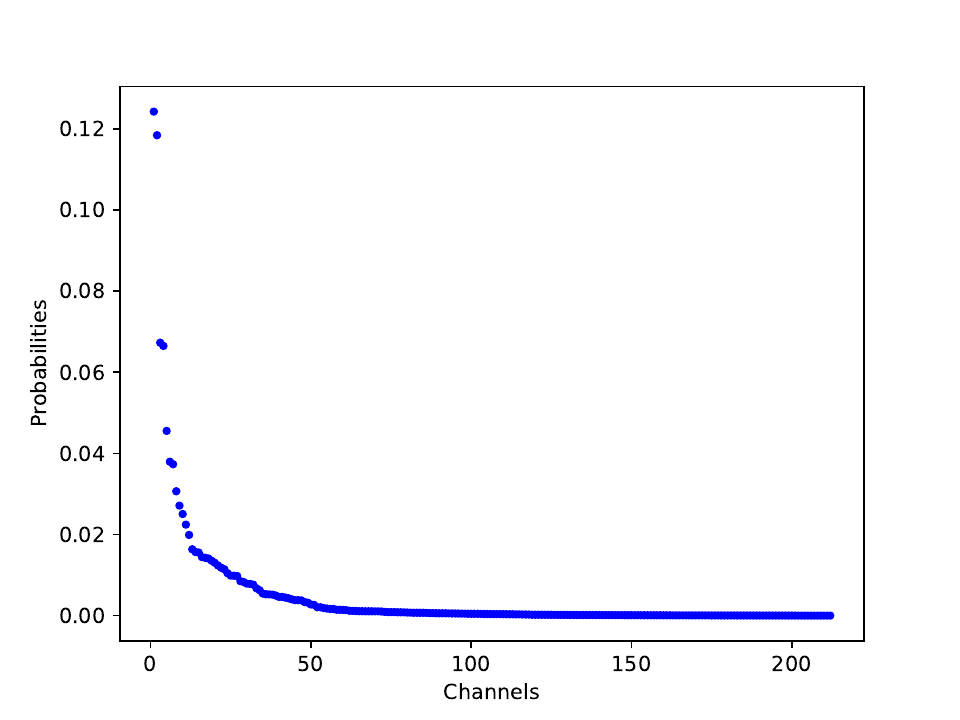}
\caption{Rank order probability distribution of ISP data}
\label{fig:ispdata}
\end{figure}

Consider a dataset that illustrates the rank order of the number of viewing channels utilized by customers over a two-hour period at a prominent European Internet Service Provider (ISP). The dataset also includes the start and end times of each viewing session, as well as the content identification number.
In this study, we focus on a specific subset of users who initiated the viewing of TV channels for a minimum of one minute within a two-hour observation window. As illustrated in Fig.~\ref{fig:ispdata} the rank order probability distribution corroborates the right-skewed nature of the ISP data, thereby substantiating the rationale for employing the LFU cache algorithm.

\begin{figure}[htp] 
    \centering
    \subfigure[LFU]
    {\label{fig: starvation}\includegraphics[width=0.48\textwidth]{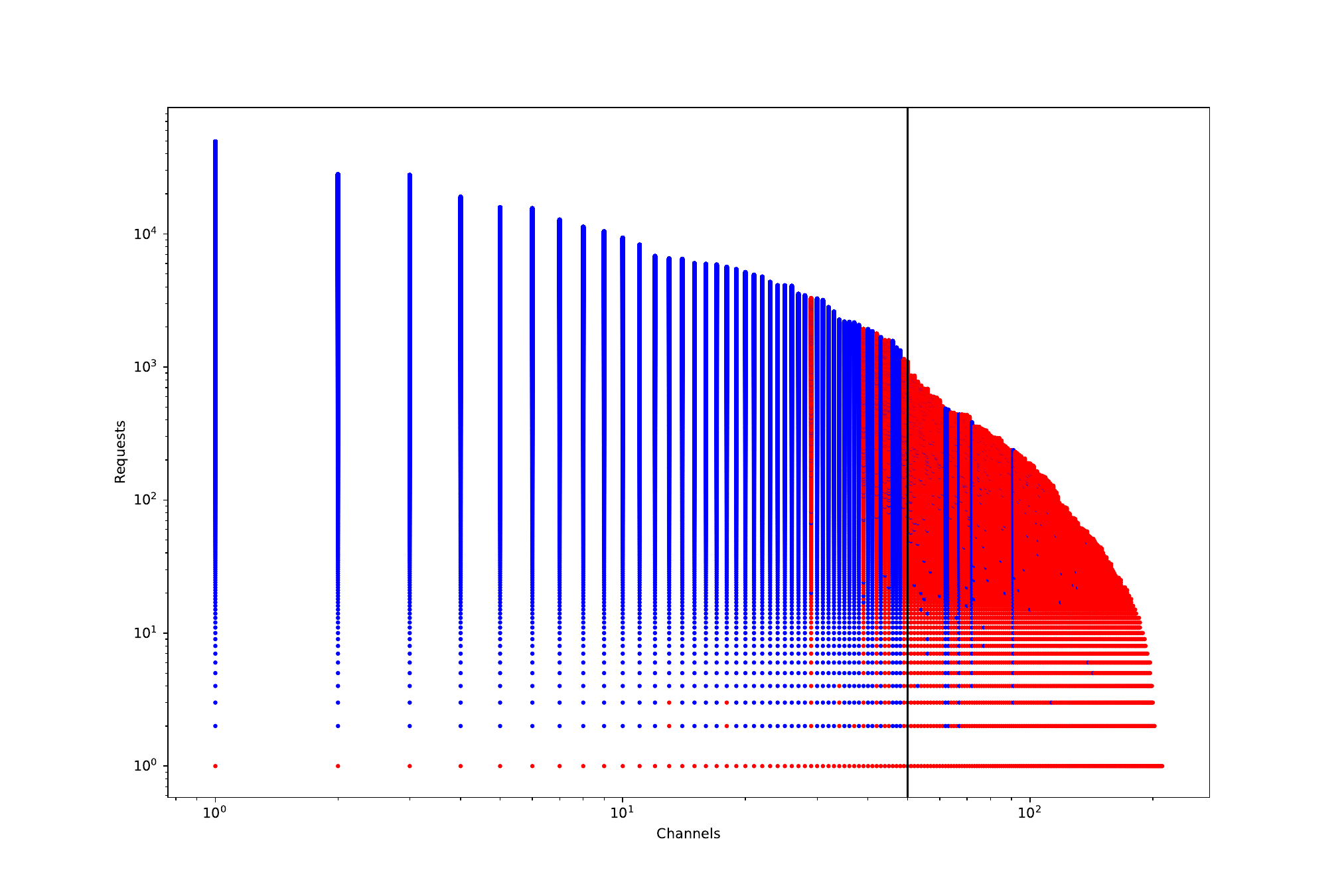}} 
    \subfigure[PLFU]
    {\label{fig: MLFU}\includegraphics[width=0.48\textwidth]
    {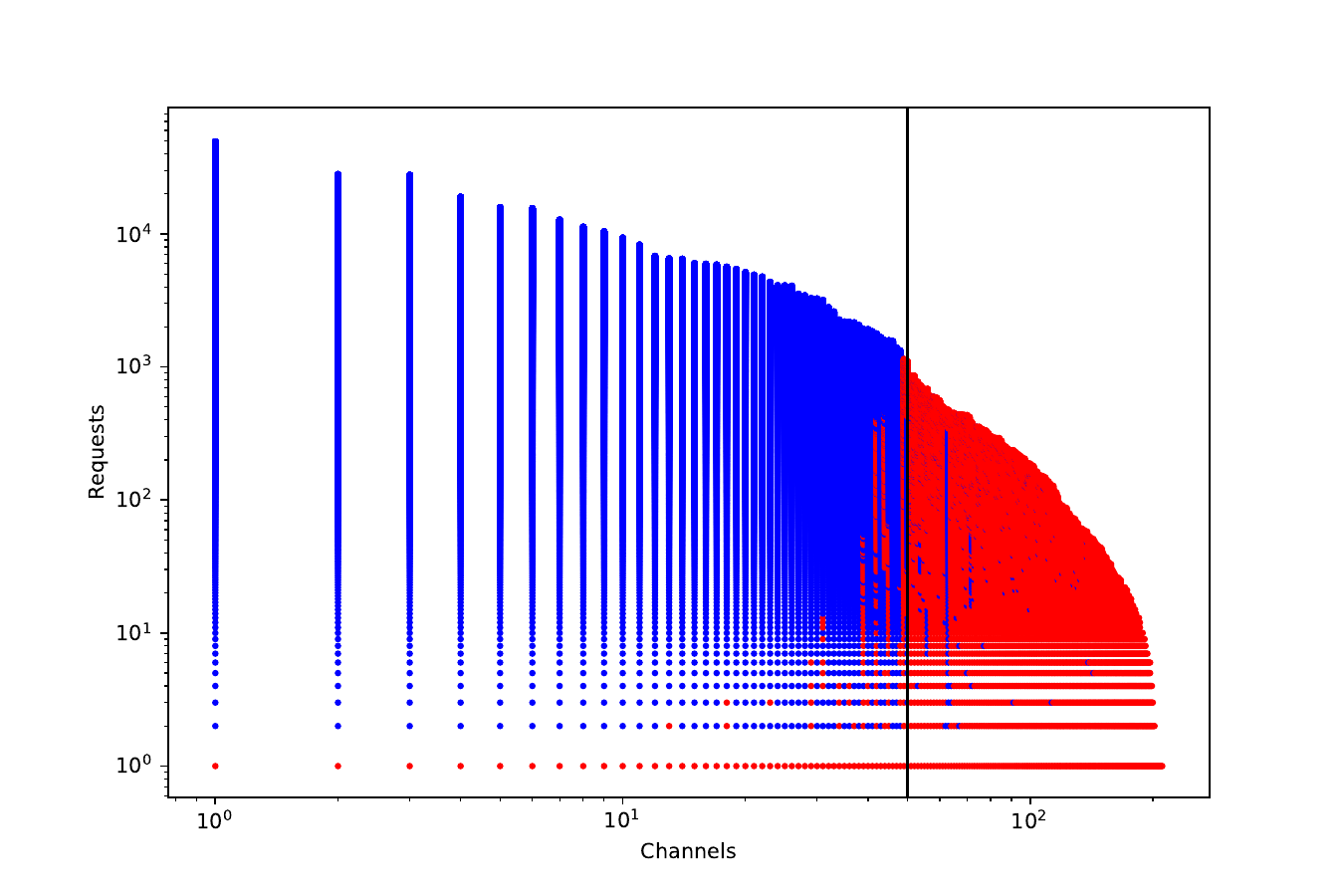}} 
    \caption{Rank order of hits and misses on a loglog scale}
    \label{fig:ispdataloglog}
\end{figure}

In order to gain insight into the performance of LFU on these channel objects, an investigation into the number of hits and misses per channel has been conducted. Fig.~\ref{fig: starvation} provides insight into the actual occurrences of hits (blue dots) and misses (red dots) while utilising the classic LFU algorithm.

The black vertical line at \(x = 50\) indicates the cache size. The discrete channel IDs are represented on the horizontal axis (212 in total), with dots above each ID in request order. The blue dots indicate hits and red dots misses for the specified channel IDs. 


From Fig.~\ref{fig:ispdataloglog}, it is evident that the initial dots above each channel are red, indicating that the inaugural call of any channel must be a miss since it is not yet in the cache. Furthermore, it can be observed that for the highly popular channels, all subsequent calls are hits. This is also logical, given that the LFU algorithm identifies these channels as having high frequencies, thus preventing their eviction from the cache. Furthermore, it is evident that to the right of the black vertical line, which represents the cache size, the majority of calls are misses (red dots). This indicates that there is no space in the cache memory for these less frequently viewed channels.

A closer examination of Fig.~\ref{fig: starvation} reveals the presence of a regular pattern within the scatter diagram. Specifically, there are red columns above certain channels on the left of the black vertical line. What is the reason for their complete or near-complete red coloration? In essence, the original LFU records the frequencies with which objects are present in the cache. Subsequently, upon a new request, the frequency of the object will recommence from 1. It is possible that an object may have been evicted with a relatively high frequency due to the fact that all other objects in the cache exhibited even higher frequencies. Following a call, the frequency resets to 1, increasing the likelihood of the cached item being evicted once more from the cache. Consequently, this process is repeated, resulting in the formation of red columns above certain channels with relatively high popularity.  

What modification of the LFU algorithm would result in a reduction of the size of the red columns that appear for objects in close proximity to the black vertical line, as illustrated in Fig.~\ref{fig: starvation}? It is evident that a novel approach to metadata storage is necessary. This is referred to as the parked-list in this paper and is utilized to document the cache frequencies of items that have previously been evicted from the cache. Upon a new request for such objects, the usual in-cache frequencies should not be reset to 1 but rather, the frequency value stored in the parked-list should be used. This solution is referred to as the Perfect LFU in the literature \cite{Breslau1998}. It has been demonstrated to outperform LFU in terms of CHR, although this is achieved at the expense of additional metadata, namely the parked-list. However, this does result in an increased memory footprint.

\subsection{PLFU algorithm on media objects} 

Fig.~\ref{fig: MLFU} demonstrates the rank order of channels and the corresponding hits (represented by blue dots) and misses (represented by red dots) above each channel object, in a manner analogous to Fig.~\ref{fig: starvation}. It is evident that the columns that were either fully or almost fully red have disappeared. However, some red columns remain, albeit shorter. A comparison of Fig.~\ref{fig: starvation} with Fig.~\ref{fig: MLFU} reveals that the full red columns have ceased to exit. In addition, a substantial number of new blue dots have resulted in a significant change to the previous pure red columns, which have now taken on a blurry reddish and bluish appearance. It is anticipated that the CHR value will typically be higher when PLFU is applied to a highly skewed distribution. Indeed, the CHR value increased from 0.9169 to 0.9349 when the PLFU algorithm was employed in lieu of the classic LFU algorithm on the ISP data set.

\subsection{Comparison of CHR values for PLFU and LFU on Zipf data}\label{CHR comparisons}

It is evident that the Cache Hit Ratio (CHR) can be employed as a performance metric for cache algorithms from an energy efficiency standpoint. This is because for cache algorithms exhibiting a high CHR, the energy saving objectives are likely to have been achieved, given that the load on the network is comparatively reduced in comparison to algorithms providing a low CHR.

As evidenced by studies such as \cite{Ntougias2021}, \cite{Einziger2017}, \cite{Breslau1998}, \cite{Shi2005} and \cite{Karakostas2000} Zipf-distributed requests are regarded as a realistic benchmark for accessing web objects. The Zipf(1.1) distribution model was selected in this paper because it provided the best fit for the observed ISP data as illustrated in Fig.~\ref{fig:ispdata}.

\begin{figure}[htp] 
    \centering
    \subfigure[LFU]{\label{fig:LFU_CHR}\includegraphics[width=0.48\textwidth]{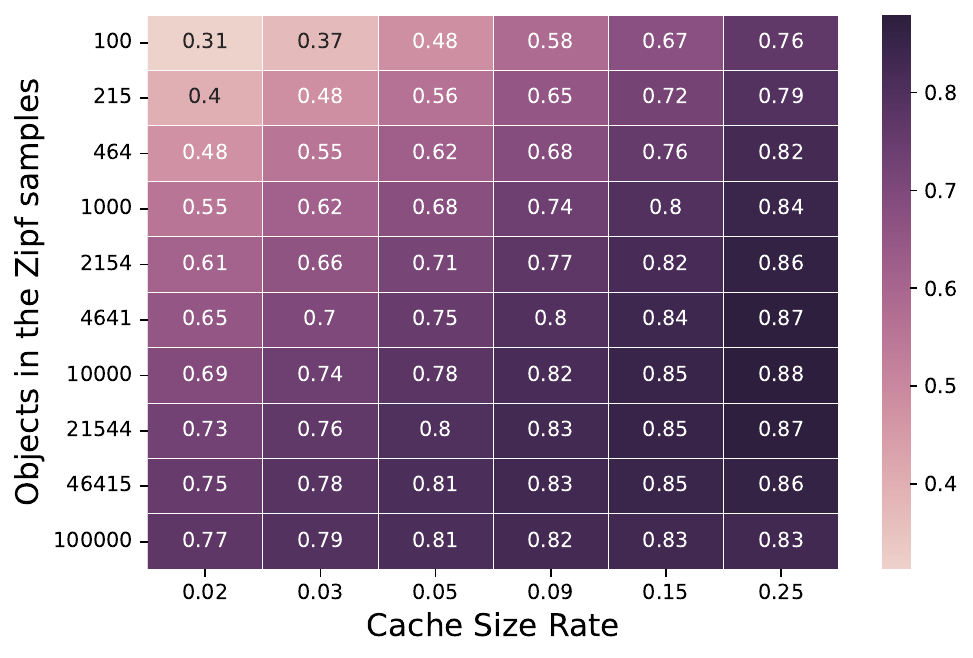}} 
    \hfill
    \subfigure[PLFU]{\label{fig:PLFU_CHR}\includegraphics[width=0.48\textwidth]{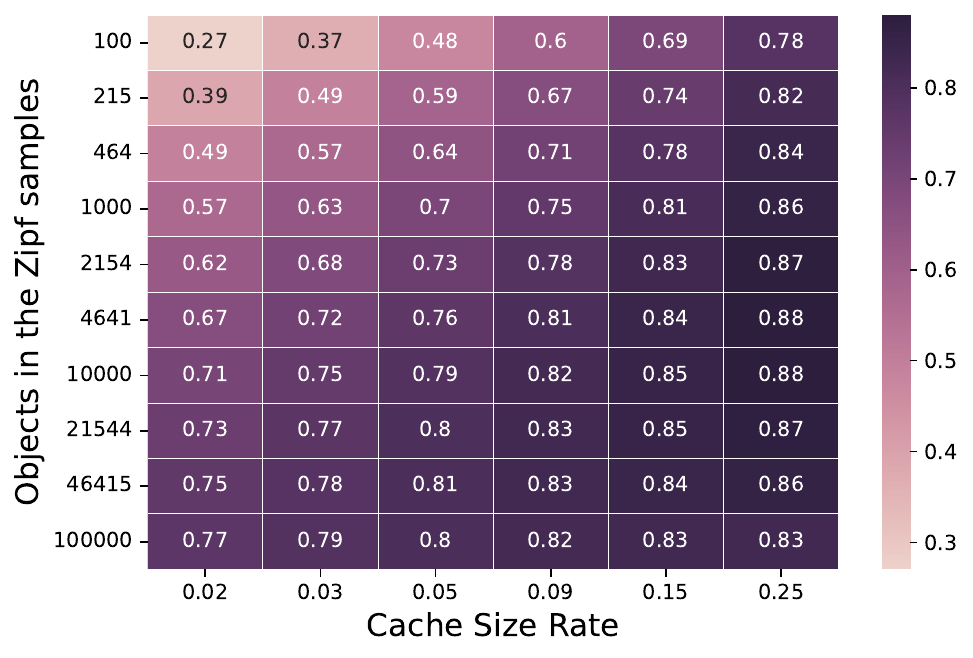}} 
     \caption{CHR values obtained from 12 random sample }
    \label{fig:CHRs}
\end{figure}

The number of objects in our samples varies between 100 and 100,000 spaced evenly on log scale. The cache sizes employed may be calculated by multiplying the rate by the number of objects. Furthermore, the employed rates also vary evenly on a log scale between 2 and 25\% as illustrated in Fig.~\ref{fig:LFU_CHR} and Fig.~\ref{fig:PLFU_CHR}. A total of 60 distinct cases were examined, with 12 random samples of 100,000 data point generated for each case, following Zipf(1.1) distribution. Fig.~\ref{fig:LFU_CHR} and Fig.~\ref{fig:PLFU_CHR} present the mean of CHR values for LFU and PLFU algorithms. 

Fig.~\ref{fig:CHRs} demonstrates a positive correlation between cache size and CHR value, with larger cache sizes yielding higher CHR values. Furthermore, it can be observed that the CHR values increase in conjunction with the number of objects present in the samples, with the exception of the bottom right corners, where both cache sizes and the number of objects are high. In this region, the CHRs exhibit fluctuations, with some instances of decline or even reduction, as the row index increases.


As previously stated in Section~\ref{introduction}, the CHR is not the sole metric to be considered, particularly when the used energy is a significant variable. The results indicate that an increase in cache size is associated with an improvement in the performance of the cache algorithm. How might this approach be modified if energy efficiency were also taken into account?

\section{Down-scaling the cache size may increase energy efficiency}

As previously discussed, in addition to the CHR values, the load on the CPU, which can be quantified as the total running time required for a cache algorithm, can also impact the evaluation of a cache strategy in term of performance. It is important to note that our implementations of the various cache types include only the cache management algorithms; the cache itself \emph{does not store or move any content}. Rather, it merely maintains its metadata container. This enables us to quantify the CPU time associated with the eviction algorithm exclusively, while any discrepancies resulting from CHR enhancements, such as a reduction in the number of content fetches, can be excluded from our assessment. This methodology ensures a true comparison.

The total CPU time for a given function can be obtained through the utilization of a profiling tool within the Python programming language. The cache implementation is single-threaded and does not rely on I/O, as this component has been excluded from the cache. Consequently, the lower-overhead cProfile profiler in Python has been applied. In light of the known limitations of the profiler, we measured the total time of a loop in our cache implementation, which enabled us to increase the measured time to a range that was multiple magnitudes greater than the resolution of the profiler's internal timer. Furthermore, it was assumed that calibration was unnecessary since the event handling time was considerably shorter than that of the various function calls. Consequently, it was deemed that any cumulative errors would be insignificant and would not impact the results.

The option of isolating CPU cores (by restricting the Linux kernel, all services and interrupts to specific cores) and running the measurement on a truly idle CPU core was also considered. However, due to the simplicity of the cache algorithm, we obtained reliable results without implementing it.

The same Zipf samples as in Section \ref{CHR comparisons} are employed and the mean total CPU times are calculated for each cache size rate, and object number in a manner analogous to that used for the CHR values in Section \ref{CHR comparisons}.

\subsection{Comparison of the total CPU time for PLFU and LFU on Zipf data} 

\begin{figure}[htp]
    \centering
    \subfigure[LFU]{\label{fig:CPU_LFU}\includegraphics[width=0.48\textwidth]{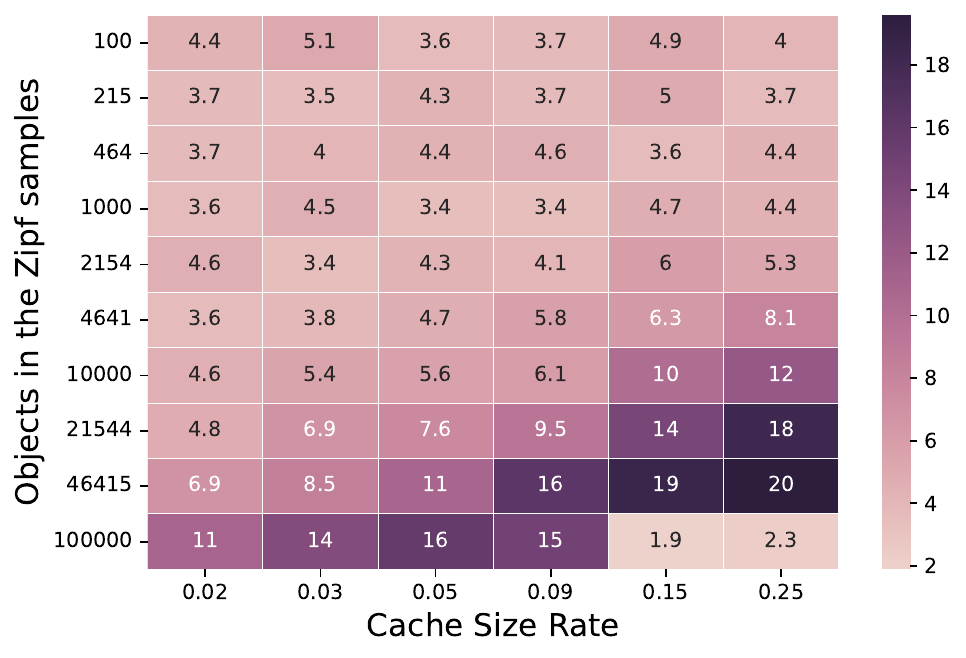}}
    \hfill
    \subfigure[PLFU]{\label{fig:CPU_PLFU}\includegraphics[width=0.48\textwidth]{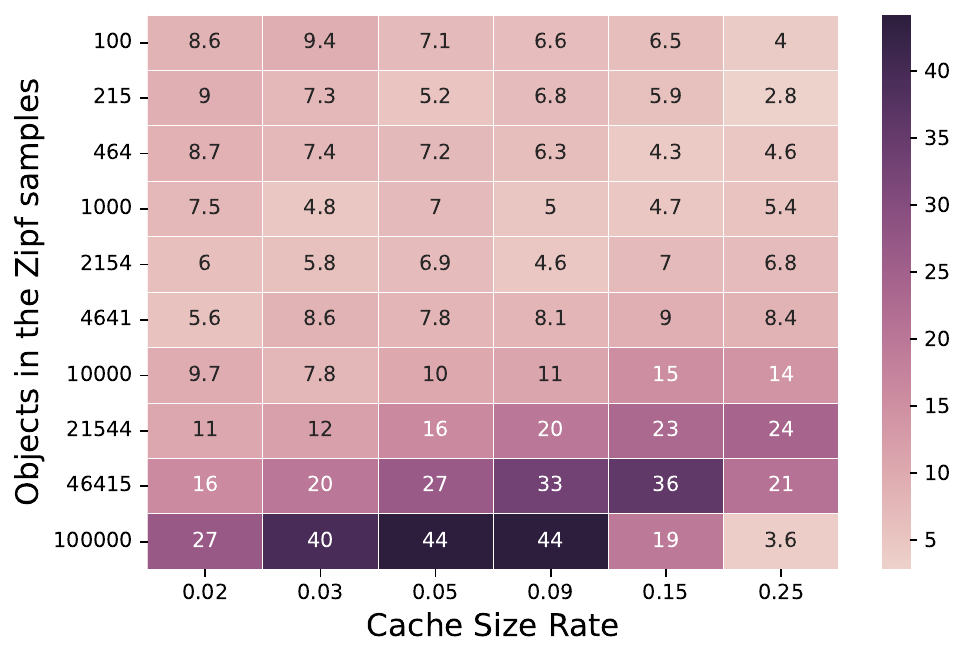}}
    \subfigure[PLFU/LFU difference]{\label{fig:CPU_incr}\includegraphics[width=0.48\textwidth]{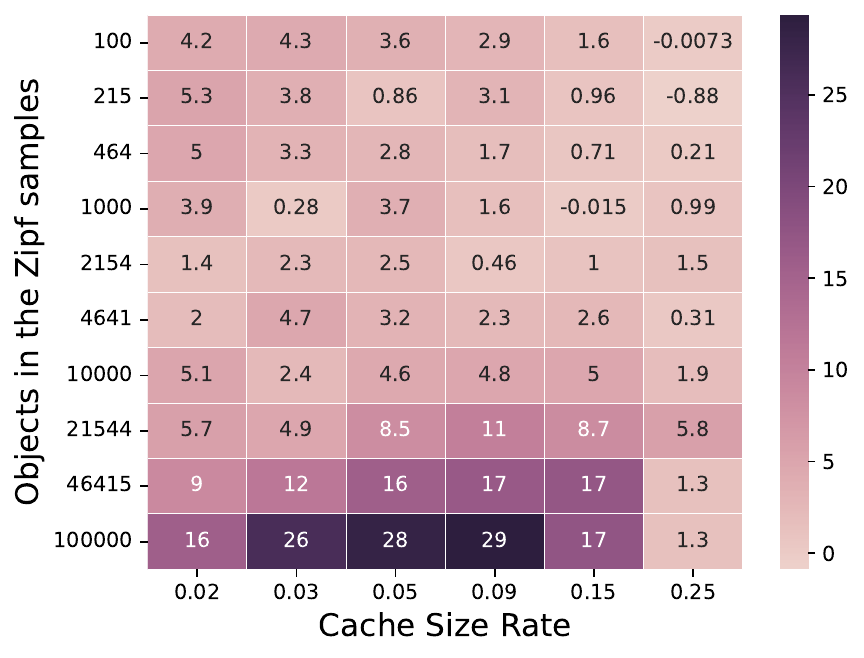}} 
    \caption{The CPU total time required}
    \label{fig:CPUs}
\end{figure}

Fig.~\ref{fig:CPUs} shows the heat-maps generated for the aggregate CPU times utilized on LFU and PLFU. It is evident that the CPU time required is typically greater for the Perfect LFU algorithm (Fig.~\ref{fig:CPU_PLFU}) than for the classic LFU (Fig.~\ref{fig:CPU_LFU}). Fig.~\ref{fig:CPU_incr} confirms that the average runtime of the Perfect LFU algorithm is significantly higher than that of the classic LFU algorithm. It is evident that the additional CPU time required for the Perfect LFU algorithm is a consequence of the increased workload associated with maintaining a secondary container for the storage and updating of request frequencies.
\renewcommand{\floatpagefraction}{.9}%

Furthermore, the same peak patterns are discernible in the total CPU time for both the PLFU and LFU algorithms. It can be observed that the right diagonals, or sharp ridges are present in the lower part of the heat matrices. This indicates that the cache size rate is to some extent independent of the CPU time. The longest CPU operation necessitated in the vicinity of the cache size of 5000 for LFU and approximately 9000 for PLFU. This can be attributed to two factors. When the cache size is small (see Fig.~\ref{fig:CPU_LFU}, row 46415, cache size 2\%), we are in the very steep part of the Zipf's PMF (probability mass function), which ensures a very clear and constant rank order of contents around the cache size. Consequently, the eviction algorithms will evict only a small number of fixed objects. If the cache size is large (see Fig.~\ref{fig:CPU_LFU}, row 46415, cache size 25\%), then we are on the long-tail part of the PMF. In this case, the majority of content is already in the cache, there is not a significant amount of content to evict. However, if we are situated somewhere between these two extremes (see Fig.~\ref{fig:CPU_LFU}, row 46415, cache size 15\%), we observe a considerable number of eviction requests, while simultaneously occupying a position on a flat part of the PMF. This results in a high number of fluctuating contents requiring eviction, which necessitates a significant amount of CPU time. The level of yield achieved is contingent upon the number of objects distributed according to the Zipf model, specifically the shape of the PMF. 

\subsection{Re-evaluating caching strategies based on metrics CHR and the used energy}

The choice of cache strategy should be based on the number of objects present in the sample. When the number of objects exceeds one hundred thousand, the LFU algorithm is observed to outperform PLFU, due to the substantial reduction in the generation of metadata. With regard to the selection of the cache size, a rate of 9\% or fifteen thousand (whichever is higher) appears to be a suitable option. The selected cache size will result in a sufficiently high CHR with a relatively short CPU time.
In instances where the number of objects ranges between one thousand and one hundred thousand, the PLFU is preferable to the LFU due to the minimal size of the generated metadata and the higher CHRs.

Furthermore, Fig.~\ref{fig:CPUs} illustrates that down-scaling the cache size is an optimal strategy for enhancing energy efficiency, as it significantly minimizes CPU utilization while maintaining a modest decline in CHR values. In particular, the recommended cache size ratio is between 5 and 9 percent in such instances.

Finally, in case the objects are below 1000 then PLFU is the optimal choice. It is not advisable to downscale, given that both the CPU operation time and the CHR values are effective for cache size rates of 0.15 and 0.25. 

\section{Perfect LFU algorithm with admission policy (PLFUA)}

Given that requests for web and media content are subject to Zipf's law, particularly the 80:20 or 90:10 rule \cite{Hasslinger2014}, it is reasonable to differentiate between the most and least popular items.

Independent Reference Model (IRS) assumes a prior probability distribution of the objects, $\ p_1 \geq p_2 \geq p_3 \geq ... \geq p_N $\ This is in accordance with the findings of \cite{Ntougias2021} and \cite{Breslau1998}. The aforementioned prior probabilities are typically obtained and undergo constant updating throughout the course of the cache operation. 

PLFUA also requires some prior knowledge about the distribution of objects, but it assumes that a group of items are designated as hot objects. In an optimal scenario, the number of hot objects would be equal to the cache size. In light of the aforementioned prior knowledge, all dots on Fig.~\ref{fig:ispdataloglog} would be colored blue (hits) to the left of the black vertical line (cache size) since only the hot objects are admitted to the cache. In this paper, however, we permit a less exact prior knowledge of popularity and label twice as many objects as the cache size as hot. This assumption is arguably realistic, given the high concentration of data. In other words, the prerequisites for PLFUA are feasible if the cache size is not large, more particularly if it less than 10\% of tthe total number of objects (N). 

In terms of the size of the metadata, PLFUA outperforms PLFU. As with the preceding section, the cache sizes obtained as proportions of the number of objects (N) between 2\% and 25\% are evenly distributed on a logarithmic scale. This indicates that the admission policy results in an admission rate of as low as 4-50\% of all objects in the discussed cases. Therefore, the size of the generated metadata is also only 4-50\% of that produced by PLFU.


\subsection{PLFUA performance evaluation}

\begin{figure}[htp] 
    \centering
    \subfigure[CHR]{\label{fig:PLFUA_CHR}\includegraphics[width=0.48\textwidth]{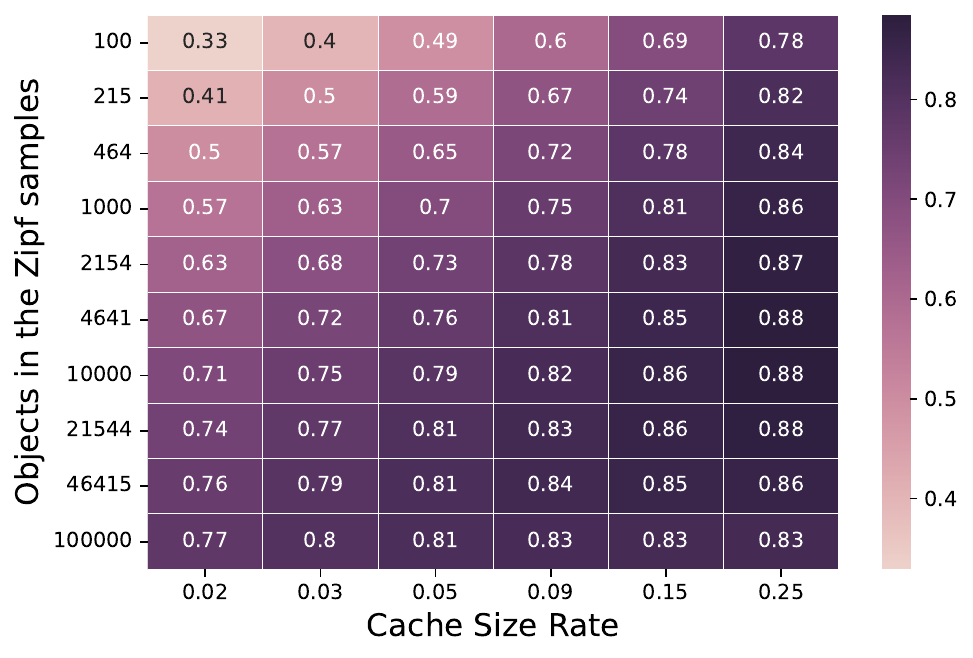}} 
    \subfigure[CPU time]{\label{fig:PLFUA_CPU}\includegraphics[width=0.48\textwidth]{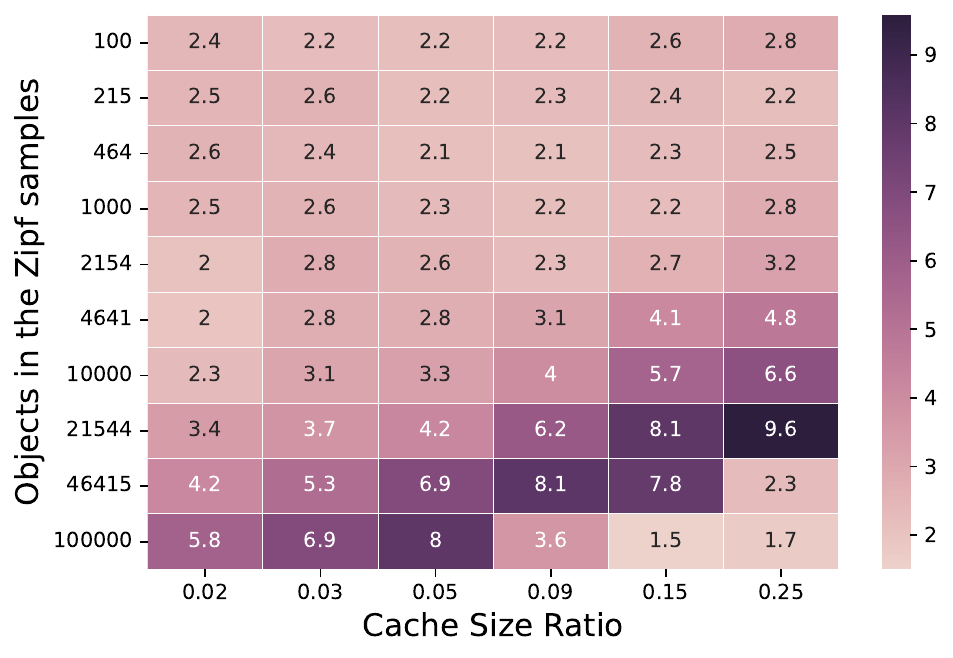}} 
    \caption{PLFUA performance metrics}
    \label{fig:PLFUA}
\end{figure}

The aforementioned cache performance metrics, which were discussed in relation to the new algorithm, are also represented by the CHR values and the total CPU execution times. These are illustrated in Fig.~\ref{fig:PLFUA}.

\begin{figure}[htp] 
\centering
\includegraphics[width=0.48\textwidth]{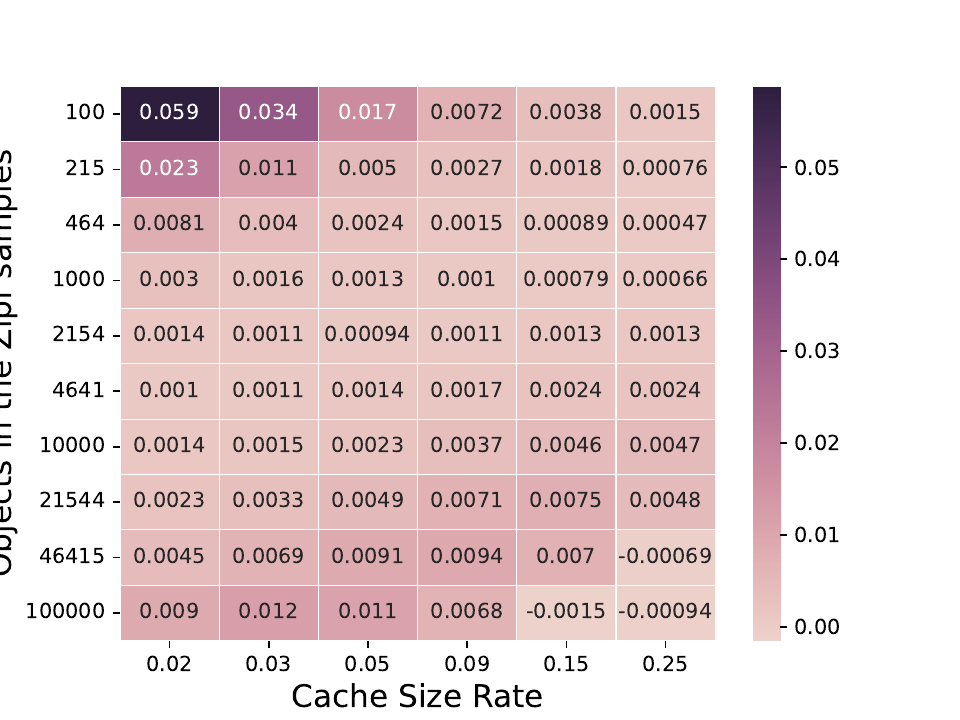}
\caption{The average increase in CHR values using PLFUA instead of PLFU} 
\label{fig: PLFUA_incr}
\end{figure}

As shown in Fig.~\ref{fig:PLFUA_CHR}, the CHR values appear to be significantly higher for cases with minimal object numbers in comparison to those observed in the PLFU. The average CHR increment for each case can be ascertained from Fig.~\ref{fig: PLFUA_incr}. With regard to the total CPU execution times, there is a notable reduction, as illustrated in Fig.~\ref{fig:PLFUA_CPU}. 

\begin{figure}[htp] 
    \centering
    \subfigure[on LFU]{\includegraphics[width=0.45\textwidth]{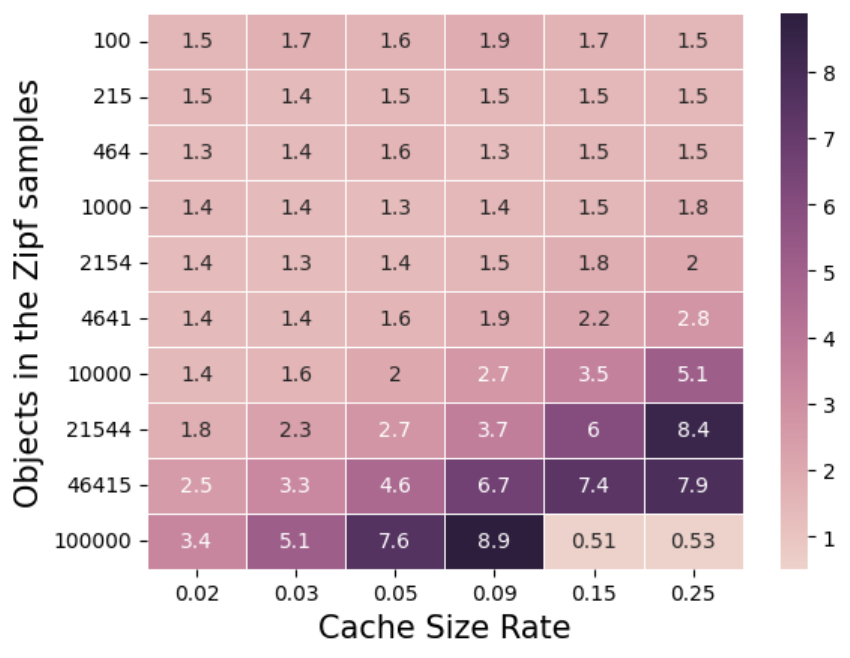}} 
    \subfigure[on PLFU]{\includegraphics[width=0.45\textwidth]{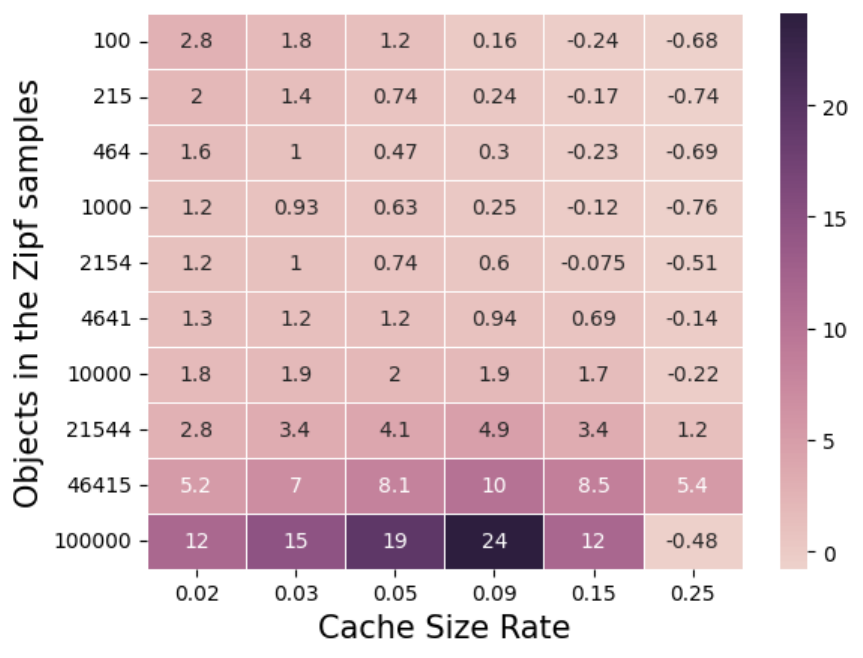}} 
    \caption{The additional CPU time required in comparison to PLFUA }
    \label{fig:extra_times}
\end{figure}

As seen in Fig.~\ref{fig:CPUs}, the execution time exhibited a notable increase in nearly all instances when PLFU was employed in lieu of LFU. Consequently, the CPU times for PLFUA were compared to those for both LFU and PLFU, as illustrated in  Fig.~\ref{fig:extra_times}.

In summary, the newly developed cache algorithm, PLFUA, has been demonstrated to outperform PLFU, the conventional cache eviction algorithm for Zipf objects, in terms of CHR and CPU execution time. The metadata generated for PLFUA is only between 4 and 50\% of the metadata size generated by PLFU, depending on the cache size rate employed. 

\section{Conclusions}

This paper presents, a graphical demonstration of the two most common frequency-based algorithms, the LFU and the PLFU on real data as seen on Fig.~\ref{fig:ispdataloglog}. Furthermore, an issue has been identified in relation to the LFU algorithm when utilizing it on actual data. The solution presented provides evidence of the existence of the PLFU algorithm in the context of skewed datasets. On the other hand, these cache eviction algorithms have also been evaluated on Zipf samples, with a focus on the cache hit ratio and the total CPU execution time. The PLFU algorithm exhibits a higher cache hit ratio, albeit at the cost of larger metadata and longer CPU time.

An intriguing consequence of our investigation is the potential benefit of smaller cache sizes, as evidenced by the satisfactory performance in terms of CHR and CPU total time.
Indeed, the sharp ridge observed on Fig. \ref{fig:CPUs} indicates that moving to smaller or larger cache sizes could result in reduced CPU execution times, albeit at the cost of a diminished cache hit ratio or an augmented cache storage size, respectively.
One significant implication of this finding is that it could be exploited by Content Delivery Networks (CDNs). In multi-tier CDNs, the selection of cache sizes is flexible to a certain extent, in that the lower CHR resulting from a smaller cache in the edge tier can be offset by larger, but fewer, acquirer caches. This enables the optimization of the overall CPU time in a CDN system.

Finally, a novel algorithm has been devised, based on PLFU, which incorporates an admission policy. The algorithm, named PLFUA, is straightforward to implement and has demonstrated superior performance compared to PLFU in terms of CHR, execution time, and metadata on Zipf samples.

\bibliographystyle{splncs04}

\bibliography{export}

\begin{thebibliography}{10}
\providecommand{\url}[1]{\texttt{#1}}
\providecommand{\urlprefix}{URL }
\providecommand{\doi}[1]{https://doi.org/#1}

\bibitem{EUROstat}
Eurostat homepage. \url{https://ec.europa.eu/eurostat/databrowser/product/page/TEN00117}, accessed: 2024-11-05

\bibitem{Breslau1998}
Breslau, L., Cao, P., Fan, L., Phillips, G., Shenker, S.: On the implications of zipf's law for web caching (1998)

\bibitem{Breslau1999}
Breslau, L., Cao, P., Fan, L., Phillips, G., Shenker, S.: Web caching and zipf-like distributions: Evidence and implications (1999)

\bibitem{Einziger2017}
Einziger, G., Friedman, R., Manes, B.: Tinylfu: A highly efficient cache admission policy. ACM Transactions on Storage (ToS)  \textbf{13},  1--31 (2017)

\bibitem{ElEnin2011}
ElEnin, S.A., ElSoud, M.A.: Evaluation of matrix multiplication on an mpi cluster. Faculty of computers and Information, Mansoura University, Egypt  (2011)

\bibitem{Hasslinger2014}
Hasslinger, G., Ntougias, K.: Evaluation of caching strategies based on access statistics of past requests. In: Measurement, Modelling, and Evaluation of Computing Systems and Dependability and Fault Tolerance: 17th International GI/ITG Conference, MMB \& DFT 2014, Bamberg, Germany, March 17-19, 2014. Proceedings 17. pp. 120--135. Springer (2014)

\bibitem{Hasslinger2016}
Hasslinger, G., Ntougias, K., Hasslinger, F.: Performance and precision of web caching simulations including a random generator for zipf request pattern. In: Remke, A., Haverkort, B.R. (eds.) Measurement, Modelling and Evaluation of Dependable Computer and Communication Systems. pp. 60--76. Springer International Publishing (2016)

\bibitem{Karakostas2000}
Karakostas, G., Serpanos, D.: Practical lfu implementation for web caching. Technical Report TR-622-00  (2000)

\bibitem{Karakostas2002}
Karakostas, G., Serpanos, D.N.: Exploitation of different types of locality for web caches. In: Proceedings ISCC 2002 Seventh International Symposium on Computers and Communications. pp. 207--212. IEEE (2002)

\bibitem{Megiddo2004}
Megiddo, N., Modha, D.S.: Outperforming lru with an adaptive replacement cache algorithm. Computer  \textbf{37},  58--65 (2004)

\bibitem{nagaraj2006web}
Nagaraj, S.: Web caching and its applications, vol.~772. Springer Science \& Business Media (2006)

\bibitem{Nair2010}
Nair, T.R., Jayarekha, P.: A rank based replacement policy for multimedia server cache using zipf-like law. arXiv preprint arXiv:1003.4062  (2010)

\bibitem{Ntougias2021}
Ntougias, G.H.J.H.K., Hasslinger, F., Hohlfeld, O.: Optimum caching versus lru and lfu: Comparison and combined limited look-ahead strategies  (2021)

\bibitem{Rashid2018}
Rashid, Z.N., Sharif, K.H., Zeebaree, S.: Client/servers clustering effects on cpu execution-time, cpu usage and cpu idle depending on activities of parallel-processing-technique operations. Int. J. Sci. Technol. Res  \textbf{7},  106--111 (2018)

\bibitem{Shi2005}
Shi, L., Gu, Z., Wei, L., Shi, Y.: Quantitative analysis of zipf’s law on web cache. In: Pan, Y., Chen, D., Guo, M., Cao, J., Dongarra, J. (eds.) Parallel and Distributed Processing and Applications. pp. 845--852. Springer Berlin Heidelberg (2005)

\bibitem{sundararajan2013smart}
Sundararajan, K.T., Jones, T.M., Topham, N.P.: The smart cache: An energy-efficient cache architecture through dynamic adaptation. International Journal of Parallel Programming  \textbf{41}(2),  305--330 (2013)

\bibitem{Thomdapu2021}
Thomdapu, S.T., Katiyar, P., Rajawat, K.: Dynamic cache management in content delivery networks. Computer Networks  \textbf{187},  107822 (3 2021). \doi{10.1016/j.comnet.2021.107822}

\bibitem{Zeebaree2015}
Zeebaree, S.R., Jacksi, K.: Effects of processes forcing on cpu and total execution-time using multiprocessor shared memory system. Int. J. Comput. Eng. Res. Trends  \textbf{2},  275--279 (2015)

\end{thebibliography}

\end{document}